\documentclass[conference]{IEEEtran}
\IEEEoverridecommandlockouts
\usepackage{cite}
\usepackage{todonotes}

\usepackage{amsmath,amssymb,amsfonts}
\usepackage{algorithmic}
\usepackage{algorithm}
\usepackage{cite}
\usepackage{graphicx}
\usepackage{textcomp}
\usepackage{xcolor}
\usepackage{booktabs}
\usepackage{gensymb}
\usepackage{siunitx}
\usepackage[inline]{enumitem}
\def\BibTeX{{\rm B\kern-.05em{\sc i\kern-.025em b}\kern-.08em
    T\kern-.1667em\lower.7ex\hbox{E}\kern-.125emX}}

\begin{document}

\title{Real-Time Energy Monitoring in IoT-enabled Mobile Devices\\
\thanks{This work was financially supported in part by the Singapore National Research Foundation under its Campus for Research Excellence And Technological Enterprise (CREATE) programme. With the support of the Technische Universit{\"a}t  {\"M}unchen - Institute  for  Advanced Study, funded by the German Excellence Initiative and the European Union Seventh Framework Programme under grant agreement n\degree~291763}
}

\author{\IEEEauthorblockN{Nitin Shivaraman$^{\dagger}$, Seima Saki$^{\dagger}$, Zhiwei Liu$^{\S}$, Saravanan Ramanathan$^{\dagger}$, Arvind Easwaran$^{\S}$, Sebastian Steinhorst$^{\ddagger}$}
	\IEEEauthorblockA{$^{\dagger}$TUMCREATE, $^{\S}$Nanyang Technological University, Singapore, $^{\ddagger}$Technical University of Munich, Germany\\
		\{nitin.shivaraman, seima.saki, saravanan.ramanthan\}@tum-create.edu.sg, \{LIUZ0052,arvinde\}@ntu.edu.sg,\\ sebastian.steinhorst@tum.de}
}

\maketitle

\begin{abstract}
With rapid advancements in the Internet of Things (IoT) paradigm, electrical devices in the near future is expected to have IoT capabilities. This enables fine-grained tracking of individual energy consumption data of such devices, offering location-independent per-device billing. Thus, it is more fine-grained than the location-based metering of state-of-the-art infrastructure, which traditionally aggregates on a building or household level, defining the entity to be billed. However, such in-device energy metering is susceptible to manipulation and fraud. As a remedy, we propose a decentralized metering architecture that enables devices with IoT capabilities to measure their own energy consumption. In this architecture, the device-level consumption is additionally reported to a system-level aggregator that verifies distributed information and provides secure data storage using Blockchain, preventing data manipulation by untrusted entities. Using evaluations on an experimental testbed, we show that the proposed architecture supports device mobility and enables location-independent monitoring of energy consumption.
\end{abstract}


\section{Introduction}

The progression of Internet of Things (IoT) towards \emph{Massive IoT}~\cite{7499809} has enabled a new wave of connected devices to have heterogeneous capabilities such as low-power communication, wide coverage area, and device-level diagnostics. This paradigm shift has enabled devices to operate independently in a decentralized architecture across all verticals of ``smart" environments ranging from smart cities to smart buildings and smart devices. However, monitoring the energy consumption of devices without a fixed central meter requires a transition from system-level to device-level metering. Moreover, decentralized metering is imperative for devices capable of mobility (\emph{i.e.,} mobile devices), that can operate in different locations on the electricity grid (grid-locations).


\begin{figure}[t!]
	\centering
	\includegraphics[width=0.97\columnwidth]{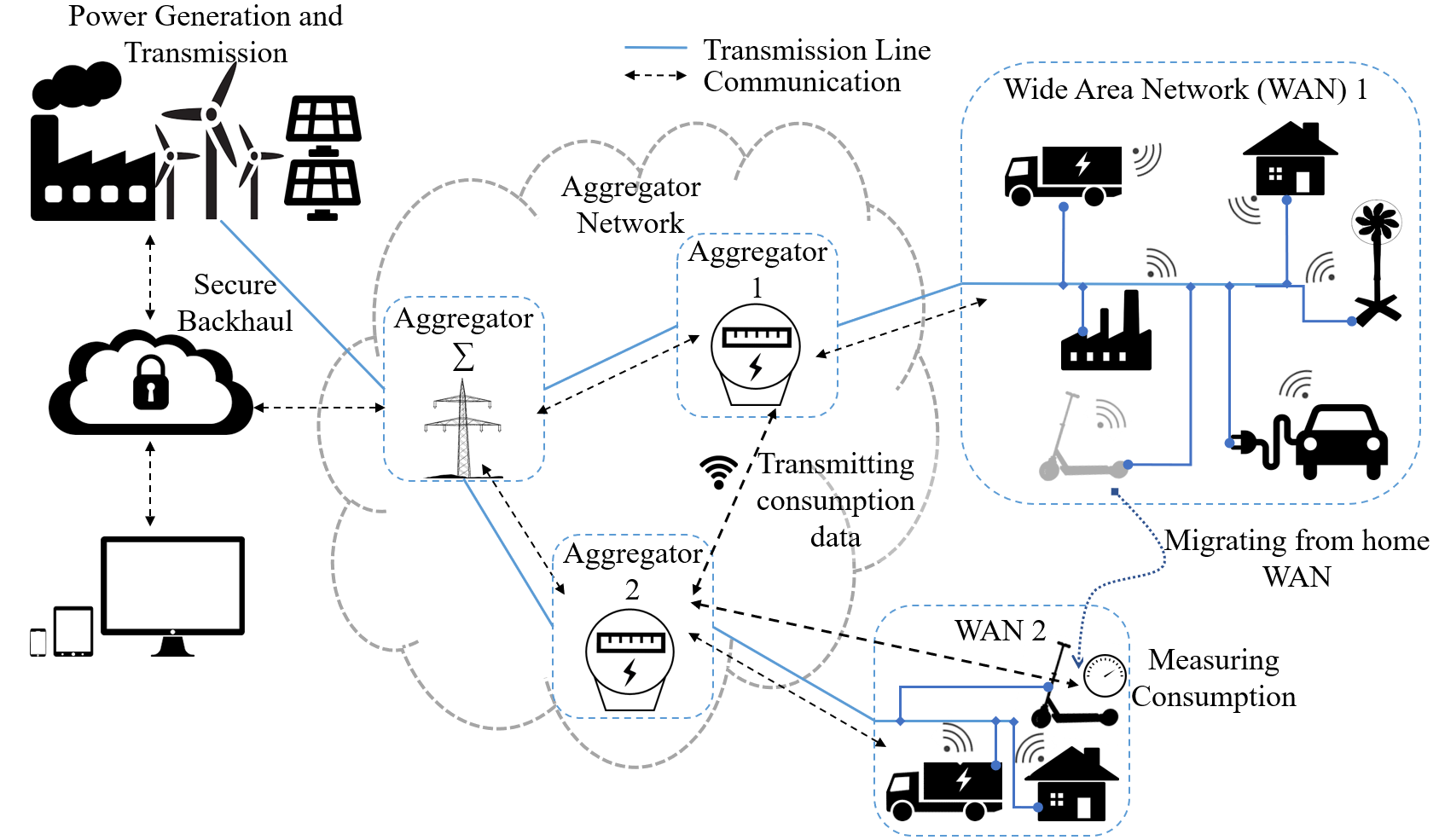}
	\caption{A decentralized real-time metering architecture.} 
	\vspace*{-5mm}
	\label{fig.prop_arch}
\end{figure}


There have been several works targeted at decentralized metering using wireless sensor networks~\cite{6102356}, energy-harvesting based custom metering hardware~\cite{6513687, 7526024} and predictive models~\cite{bayesian_inf, 7300834}. There have also been some works on detecting data tampering and secure data logging~\cite{7856180}. The relative variation in metering data combined with historical consumption data has been used to detect data tampering~\cite{8401980}. Blockchain has been used in microgrids of a decentralized energy market for secure storage~\cite{ANDONI2019143}. The methods in~\cite{6102356,6513687,7526024,bayesian_inf,7300834} are highly susceptible to data tampering (by untrusted entities) and do not provide a tamper-proof storage and the methods~\cite{6102356,6513687,7526024,bayesian_inf,7300834,7856180,8401980,ANDONI2019143} do not cater to mobile devices that can operate in multiple grid-locations. 


In this paper, we propose a decentralized metering architecture (shown in Figure~\ref{fig.prop_arch}) that enables mobile devices to monitor their energy consumption independent of their grid-location. The Wide Area Network (WAN) could represent a residential, industrial or any other communication network environment depending on the device. Let us suppose a device (say e-scooter) moves from WAN 1 (home communication network) to WAN 2 (host communication network). The e-scooter initially reports its consumption data during charging to Aggregator 1 while it stays in WAN 1 and starts reporting to Aggregator 2 once it enters WAN 2. The aggregators are interconnected through a mesh/cloud network to exchange consumption data of the devices connected to them.

Our proposed architecture has two layers of connectivity among devices: 1) physical grid representing the wired electrical connection (blue solid lines) and 2) communication network (network in short) that could be wired or wireless (black dotted lines). Each device reports its consumption to a trusted device within its home network called \emph{aggregator} for verification and hence, enables billing at the home network. The aggregator performs data aggregation of all devices within the network and acts as a nodal point for multi-hop communication with the other networks. The aggregator uses an additional system-level complementary measurement (sum, average, etc.) along with the measurements of all the devices in the network to detect anomalies in the reported value. The aggregator stores the consumption data of all the devices in the network in a blockchain. 
Since the aggregator is trusted and validates the data, there is no consensus required among devices. 
By encapsulating the consumption data into a blockchain, data storage is made tamper-proof.


The contributions of the paper are the following:
\begin{enumerate*}[label=(\roman*), itemjoin={{. }}, itemjoin*={{. }}]
	\item We propose a decentralized metering architecture capable of device-level energy metering
	\item We develop an algorithm for network transition (change of grid-location) to support mobility
	\item We demonstrate the decentralized metering with mobility through experimental evaluations on a testbed.
\end{enumerate*}
\section{Architecture}



The physical grid infrastructure comprises devices, aggregator units connected in a backhaul network, and the power generation and transmission units (see Figure~\ref{fig.prop_arch}). We assume all components remain similar to the existing grid infrastructure, except for the devices and aggregator units. Hence, we only present the details of devices and aggregator units.

\subsection{Aggregator Architecture}
\label{sec:system_architecture}
The aggregator units are deployed across different grid-locations. All the devices that are physically connected (through a transmission line) to an aggregator form a network (denoted as WAN in Figure~\ref{fig.prop_arch}) and report their consumption data to that aggregator. 
We assume that all the devices in the network and the aggregators are time-synchronized. The aggregator provides the devices with time-slots for communication to prevent interference. With limited time-slots for communication, the number of devices connected to an aggregator is also limited. The role of an aggregator is to use its measurement to establish the ground truth\footnote{The ground truth refers to the actual device consumption value as opposed to the reported measurement value.}. 



To prevent any tampering of consumption data, the reported data and a hash are encapsulated into a blockchain data structure by the aggregator. The hash of a new block is created from the reported data and the hash of the previous block. Creating the hash is not an expensive operation, and hence, does not expend significant computation power. The blocks from all the aggregators are formed into a common permissioned blockchain. Blockchain is only used as a hashed data chain without any consensus~\cite{8343261}. This is a reasonable change as the trusted aggregators validate the reported data with the ground truth before a block is created.

In a truly decentralized network, the aggregators' role could be performed by the devices themselves having a consensus among themselves. In that case, the consumption data must be broadcast to the network and a common blockchain is formed once a consensus is achieved among them. 

\begin{figure}[t!]
	\centering
	\includegraphics[width=0.96\columnwidth, height=5.8cm]{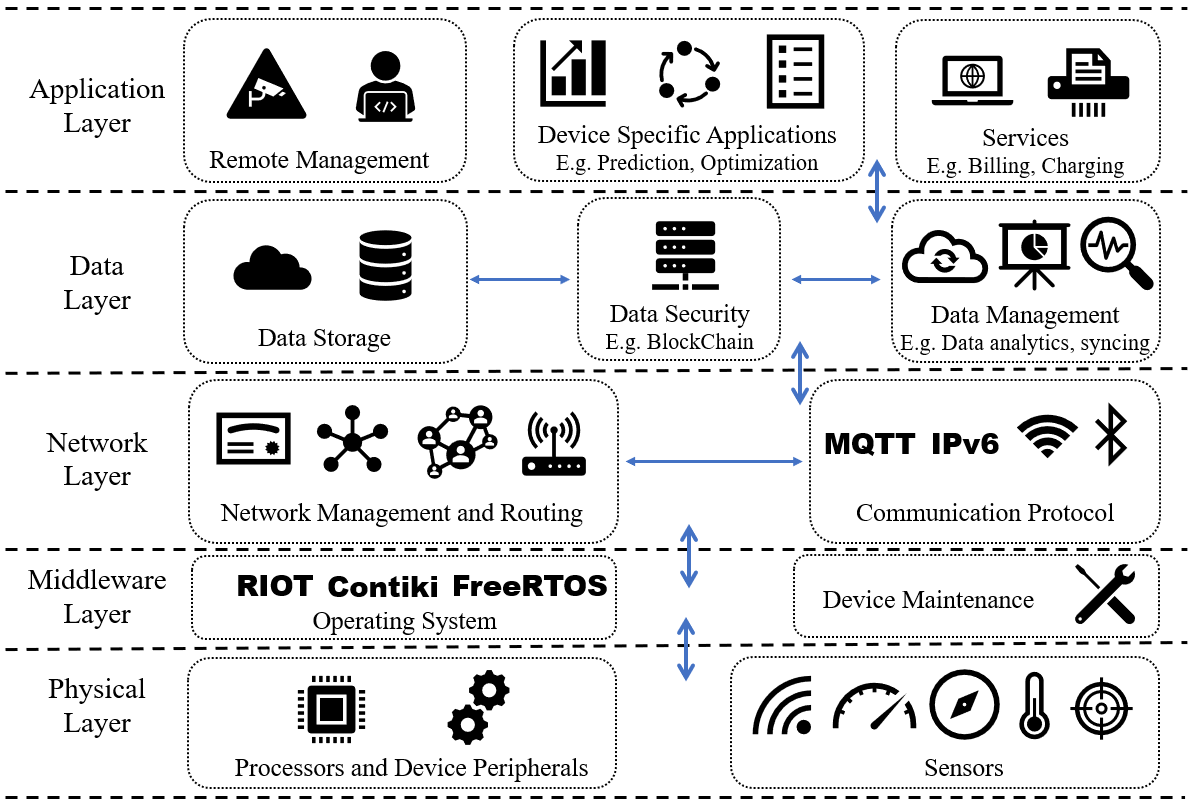}
	\caption{Device-level software architecture.} 
	\vspace*{-5mm}
	\label{fig.soft_arch}
\end{figure}

\subsection{Device Architecture}
\label{sec:device_architecture}
Each device in our architecture must be enabled with IoT capabilities to implement location-independent device metering. The software architecture of such a device with different layers is shown in Figure~\ref{fig.soft_arch}. 

The bottom-most layer is the physical layer which comprises processors, device peripherals and sensors. This layer is responsible for physical connectivity, transmission of raw data in the form of signals (electric/radio) and measurement of consumption through sensors. 

The middleware layer is mainly composed of the operating system and the firmware to control the hardware peripherals. 

The network layer has two components: network management and communication protocol.
Network address translation, synchronization of devices and aggregators, network authentication and consumption data routing are handled by the network management component. Communication protocols (such as Wi-Fi, Message Queuing Telemetry Transport (MQTT), etc.) and data packet encapsulation are handled here. 

Data representation, security, and storage are the main features of the data layer. In the absence of network connectivity with the aggregator, raw consumption data is stored in the local storage until the connection is established. 

The application layer comprises: 1) remote management for monitoring/device maintenance, 2) device-specific applications such as demand prediction and schedule optimization for better load management, and 3) services such as billing.

\subsection{Device Registration and Real-time Energy Monitoring}
\label{sec:device_registration}
In this section, we illustrate the device registration process and energy monitoring of mobile devices in our architecture. A device could either be stationary or mobile depending on its capabilities and tasks. Every device must be registered to an aggregator network irrespective of its type. A stationary device undergoes a single registration process in its lifetime while mobile devices need to register every time they migrate to a network different from their home network.

\begin{figure}[t!]
	\centering
	\includegraphics[width=0.97\columnwidth]{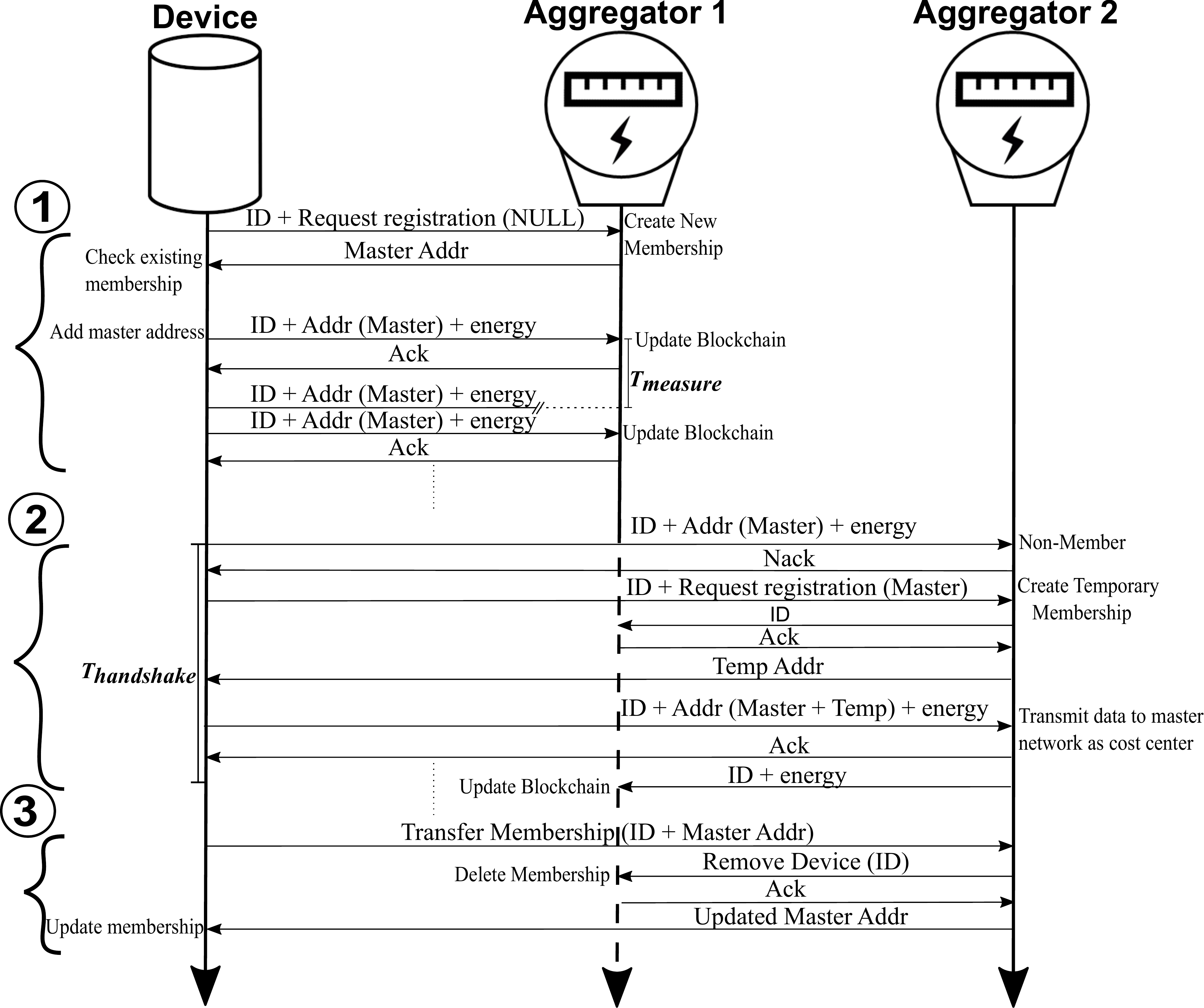}
	\caption{Device registration and real-time energy monitoring process.} 
	\vspace*{-5mm}
	\label{fig.msg_flow}
\end{figure}

\paragraph*{Membership Registration}
The membership registration process is shown in Figure~\ref{fig.msg_flow} (sequence 1). Initially, the devices are not registered to any of the network aggregators. Each device broadcasts a membership registration request (includes its registered address and ID) within its network. The network aggregator\footnote{The Received Signal Strength Indicator (RSSI) is used by the device in case of a wireless communication channel to detect its reporting aggregator.}, responds to the request with the corresponding network address (\emph{Master address}) for data transfer. 
The device updates its registered network to the Master address and starts transmitting the consumption data at a pre-configured time interval $T_\mathrm{measure}$. An acknowledgment (\emph{Ack}) is sent by the aggregator for every successfully transmitted measurement from the device. 
If there is a change in the network, a transmission or a registration failure, the raw energy consumption value while charging is temporarily stored in local memory. The combination of stored data and the measurement are transmitted to the aggregator in the next transmission. 

\paragraph*{Real-Time Energy Monitoring}
To facilitate grid-location independent real-time energy monitoring, the aggregators liaise with other aggregators to receive information on devices that are currently outside their home network. 
It is important to note that the measurement of consumption and billing is only done for the duration in which devices are connected to the grid and not during the transit.

Let us suppose a device moves from Network 1 to Network 2 as shown in Figure~\ref{fig.prop_arch}. Sequence 2 in Figure~\ref{fig.msg_flow} shows the switch from Aggregator 1 to Aggregator 2. The moment the device moves out of Network 1 (home network) it gets disconnected from Aggregator 1. 
After the transition, the device measures its consumption and reports it to Aggregator 2. Aggregator 2 upon receiving the consumption data sends a negative acknowledgment (\emph{Nack}) to indicate the absence of membership.  Upon reception of a \emph{Nack}, the device re-initiates the membership sequence to obtain temporary membership with Aggregator 2 by including its Master address (i.e., address of Aggregator 1). Aggregator 2, after verifying the device ID with Aggregator 1, creates a temporary membership to collect the consumption data from the device on behalf of Aggregator 1. These values are in turn transmitted back to the home network using the \emph{Master address} of the device. The time duration required to establish a temporary membership is denoted as $T_\mathrm{handshake}$. Further, time to transmit the consumption data to Aggregator 1 is dependent on the backhaul network among aggregators. If the device moves out of Network 2, the temporary membership is immediately discarded. Based on the received data from the Aggregator 2, the home network can continue billing the device for its consumption in the external network. Meanwhile, the home network retains the membership of the device at all times unless there is a message to remove it due to loss/reset/transfer-of-ownership of the device as shown in sequence 3 of Figure~\ref{fig.msg_flow}. 

\section{Experiments}
\label{sec:experiments}

\begin{figure}[t!]
	\centering
	\includegraphics[width=0.92\columnwidth]{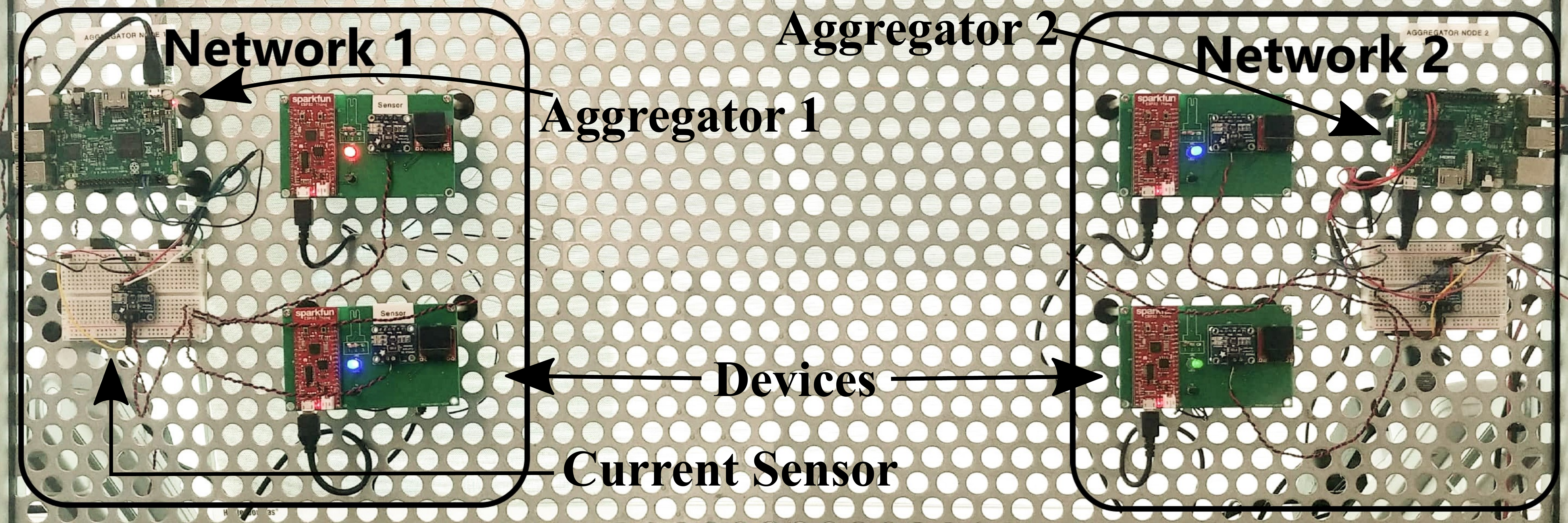}
	\caption{The experimental testbed consisting of two Aggregators, each connected to two devices.}
    \vspace*{-2mm}
	\label{fig.testbed}
\end{figure}

\subsection{Experimental setup}
\label{sec:experimental_setup}
The experimental setup consists of two networks each with two devices using Sparkfun ESP32 Thing~\cite{esp32} and an aggregator using Raspberry Pi (RPi) Model B. All the devices and aggregators are equipped with INA219~\cite{tiina219}, a current monitoring sensor and DS3231~\cite{rtcmodule}, a real-time clock module. The testbed is shown in Figure~\ref{fig.testbed}. Using the voltage characteristics of the device, the energy consumption is computed using the sensor measurement value and the measurement duration. MQTT protocol is used for consumption data transfer to the aggregator over Wi-Fi. We use Grafana~\cite{grafana} to monitor live data transmission.



\begin{figure}[t!]
	\centering
	\includegraphics[width=0.8\columnwidth, height=5.5cm]{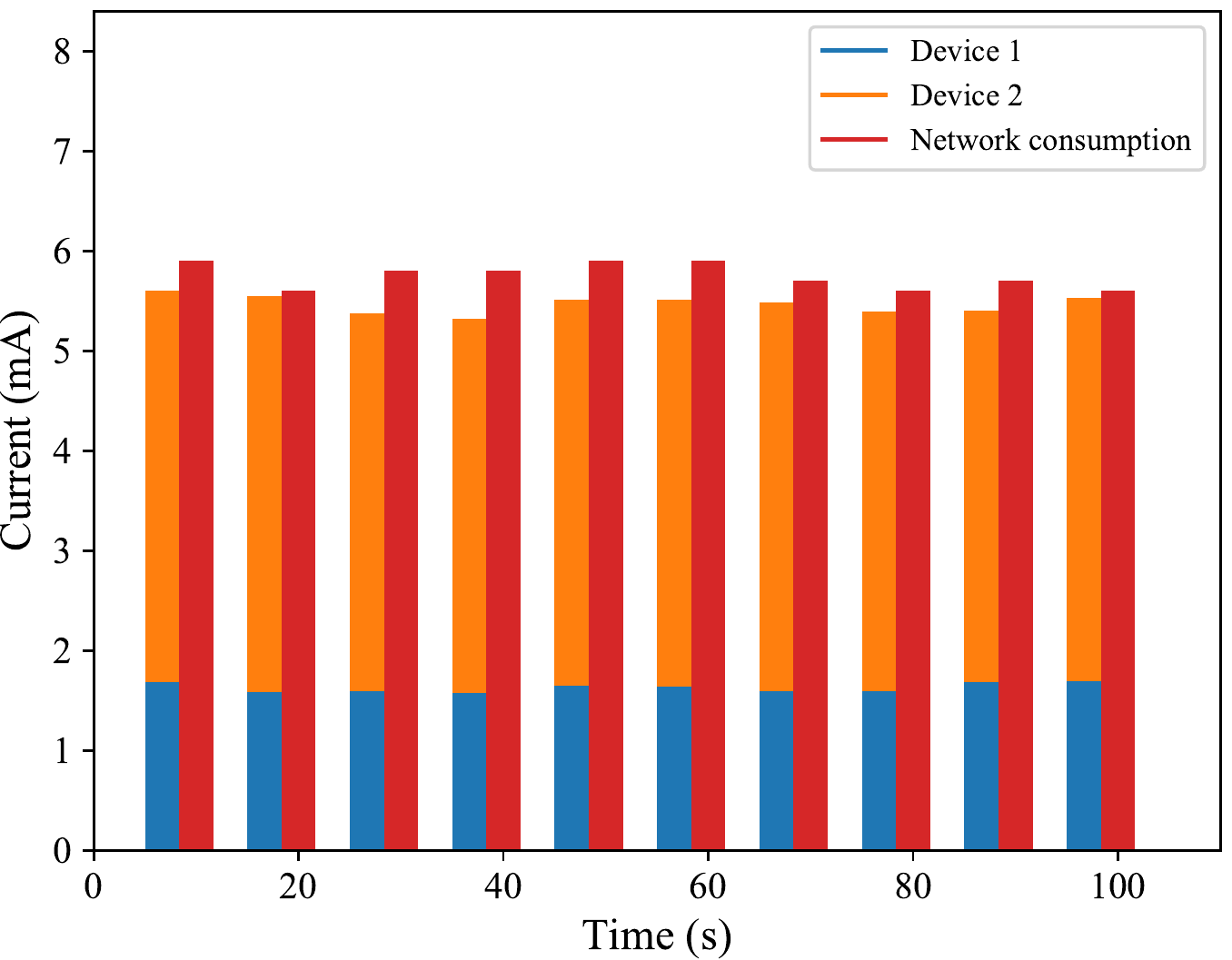}
	\caption{Comparison of individual device measurements with the network aggregator measurement.}
	\vspace*{-5mm}
	\label{fig.aggregation}
\end{figure}

\subsection{Results}
\label{sec:results}
We present the results from two experiments that were carried out to evaluate our architecture. 

\paragraph{Decentralized Metering}
The first experiment was designed to evaluate the measurement accuracy of decentralized metering over centralized metering. The aggregator in our setup has a physical electrical connection with the rest of the network and provides the total energy consumption for the network which is analogous to a centralized meter. This is compared with the aggregation of reported values of the devices. Figure~\ref{fig.aggregation} shows a stacked bar chart with the left bars indicating the individual consumption of devices and the right bars indicating the aggregator measurement of the overall consumption of the network. We observed that the consumption value at the aggregator is slightly ($0.9-8.2\%$) higher than the aggregated values as shown in Figure~\ref{fig.aggregation}.
This is due to the ohmic losses of various electrical components and the measurement error of the current sensor in the setup. The current sensor has an offset error of 0.5mA~\cite{tiina219} which contributes to the variation observed. 

\begin{figure}[t!]
	\centering
	\includegraphics[width=0.8\columnwidth, height=5.5cm]{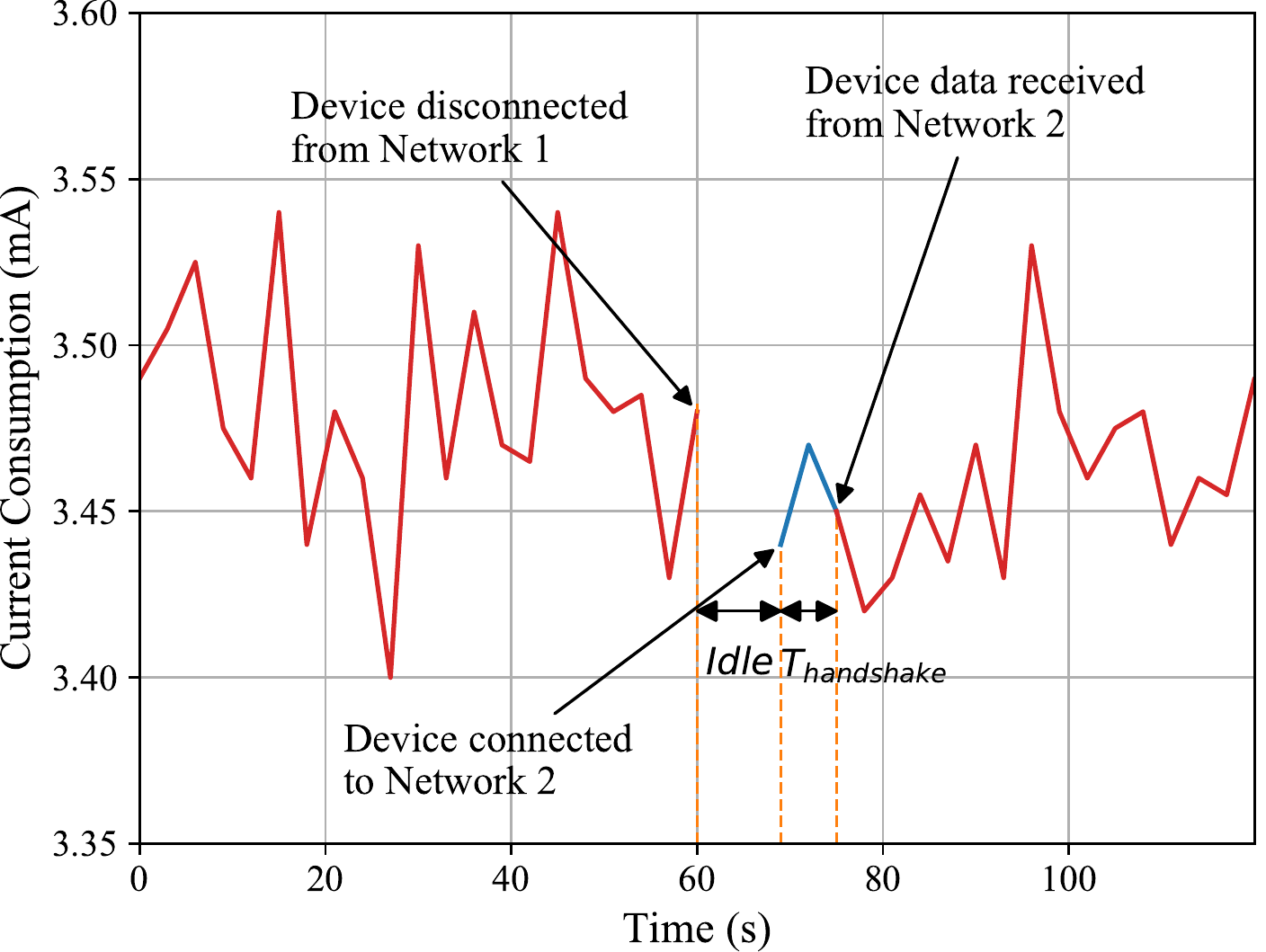}
	\caption{Current consumption reported at Aggregator 1 for a mobile device transiting from network 1 to network 2, before and after connection establishment with Aggregator 2.}
	\vspace*{-5mm}
	\label{fig.net_switch}
\end{figure}

\paragraph{Device Mobility}
The primary benefit of our architecture is device mobility which aids in per-device billing. Our second experiment is hence designed to validate the claim that the device consumption can be monitored even when it is operated at different grid-locations. To demonstrate this, we move one of the devices from one network (Network 1) to another network (Network 2). The consumption data of the device during the network transition obtained from Aggregator 1 is shown in Figure~\ref{fig.net_switch}. The device is initially registered to Network 1 and reports its consumption value to Aggregator 1. The reported values (until it gets disconnected from Network 1) are shown in the left half of the figure. The pre-configured measurement interval for the device, $T_\mathrm{measure}$, was set to 10 times per second i.e.,  the device consumption is reported to the aggregator every \SI{100} {milliseconds}. If the device is disconnected before the reporting time, the data is stored locally until the network is restored. After one hour, the device is moved from Network 1 to Network 2.


When the device is moving across different networks there is no consumption (as it is not connected to the transmission line) and this duration is denoted as \emph{Idle} time in the figure. Once the device establishes an electrical connection with a different power network it continuously scans the communication network to determine its reporting aggregator (Aggregator 2 in our case). The device stores its consumption (marked by the blue line in the figure) during the handshake process to local storage until the network connection is established. After establishing a connection (\emph{i.e.,} temporary membership registration) the device transmits consumption data and any locally stored data to Aggregator 2. Aggregator 2 communicates this data back to Aggregator 1 for consolidated billing. The time to register a temporary membership in Network 2, $T_\mathrm{handshake}$, is found to be \SI{6} {seconds} on average with a variation between \SI{5.5}{-}\SI{6.5}{seconds} over 15 runs. The data communication between aggregators does not incur much delay (\SI{1}{milli second}) as the backhaul network is assumed to have high bandwidth.

\section{Conclusion and Future Work}
\label{sec:conclusion}
In this paper, we presented an architecture for secure decentralized in-device metering for IoT-enabled devices. We presented the salient features of our architecture and illustrated the real-time energy monitoring process.
Through experiments, we showed the feasibility of measuring the device energy consumption outside its home network.

Although decentralized metering is a favored solution for mobile devices, initial implementation necessitates a higher investment cost as the devices need to be IoT-enabled with metering capabilities. 
Further, device mobility introduces unprecedented demand variability and leads to research problems such as dynamic load-balancing.
Addition of consensus among devices to realize a completely decentralized without any reliance on the aggregator is planned. We also plan to address the ground truth problem to identify an anomalous device that reports data different from its actual consumption.




\bibliographystyle{IEEEtran}
\bibliography{references}
\vspace{12pt}

\end{document}